\documentclass[aps,pra,reprint,amsmath,amssymb,superscriptaddress,showkeys]{revtex4-1}

\usepackage{graphicx}
\usepackage{dcolumn}
\usepackage{bm}
\usepackage{xcolor}
\usepackage{makecell}
\usepackage{ctable}

\begin{document}

\title{Evolving reservoir computers reveals bidirectional coupling between predictive power and emergent dynamics}

\author{Hanna M. Tolle}
\email{h.tolle23@imperial.ac.uk}
\affiliation{Department of Computing, Imperial College London}

\author{Andrea I Luppi}
\affiliation{Department of Psychiatry, University of Oxford}
\affiliation{Division of Information Engineering and St John’s College, University of Cambridge}

\author{Anil K. Seth}
\affiliation{Department of Informatics, University of Sussex}
\affiliation{Sussex Centre for Consciousness Science, University of Sussex}

\author{Pedro A. M. Mediano}
\affiliation{Department of Computing, Imperial College London}
\affiliation{Division of Psychology and Language Sciences, University College London}

\begin{abstract}
Biological neural networks can perform complex computations to predict their environment, far above the limited predictive capabilities of individual neurons. While conventional approaches to understanding these computations often focus on isolating the contributions of single neurons, here we argue that a deeper understanding requires considering emergent dynamics -- dynamics that make the whole system ``more than the sum of its parts''. Specifically, we examine the relationship between prediction performance and emergence by leveraging recent quantitative metrics of emergence, derived from Partial Information Decomposition, and by modelling the prediction of environmental dynamics in a bio-inspired computational framework known as reservoir computing. Notably, we reveal a bidirectional coupling between prediction performance and emergence, which generalises across task environments and reservoir network topologies, and is recapitulated by three key results: 1) Optimising hyperparameters for performance enhances emergent dynamics, and vice versa; 2) Emergent dynamics represent a near sufficient criterion for prediction success in all task environments, and an almost necessary criterion in most environments; 3) Training reservoir computers on larger datasets results in stronger emergent dynamics, which contain task-relevant information crucial for performance. Overall, our study points to a pivotal role of emergence in facilitating environmental predictions in a bio-inspired computational architecture.

\end{abstract}

                              
\maketitle

\section{\label{sec:Introduction}Introduction}

Biological nervous systems may be regarded as collections of weak learners: individual neurons, each with limited predictive power over the system's environment. Yet, as a whole, these networks perform highly non-trivial computations that support the prediction of environmental dynamics necessary for survival.

From a reductionist perspective, a complete understanding of a system's computations can in theory be achieved by describing the behaviour of each of its components and their interactions. Yet in practice, it is overwhelmingly the case that biological neural networks are partially observable systems, rendering a full description of components and their interactions intractable.

\begin{figure*}[t]
    \centering
    \includegraphics{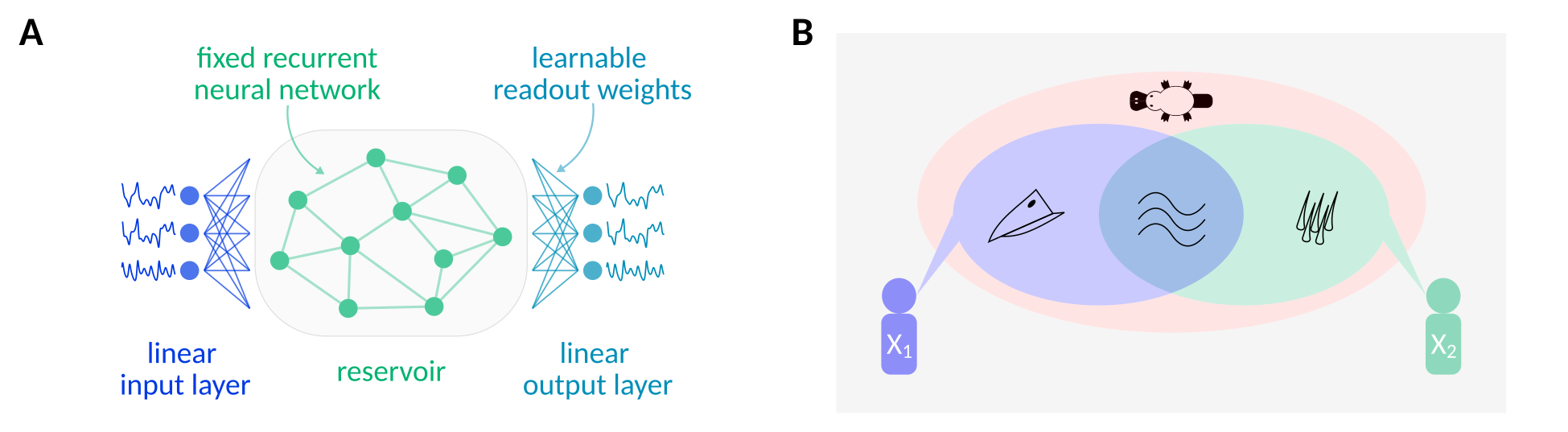}
    \caption{\textbf{Schematic illustration of central methodological concepts.} (\textbf{A}) The basic components of a reservoir computer. The reservoir, a recurrent neural network with fixed internal weights, serves as the core processing unit. Inputs are fed to the reservoir via a linear weight matrix. Outputs are computed by another linear weight matrix, the so-called readout weights, which are learned. (\textbf{B}) Illustrative example of the partial information decomposition (PID) of Shannon's mutual information into unique, redundant and synergistic information. Two witnesses ($X_{1}$ and $X_{2}$) are interrogated to identify the type of animal ($Y$) that seems to have consumed large parts of the seasonal harvest. $X_{1}$ reports that the animal had a beak and when they approached, it fled to hide in the pond. $X_{2}$ confirms that they also observed the animal swimming in the pond and additionally, they saw that the animal was furry. Taken together, both sources $X_{1}$ and $X_{2}$ provided redundant information (that the animal can swim) and unique information (that the animal has a beak and is furry) about the target $Y$. However, the identity of $Y$ was only revealed after the information from both $X_{1}$ and $X_{2}$ was combined (synergistic information), because the only animal with a beak and fur is a platypus!}
    \label{fig:introFigure}
\end{figure*}

In this context, machine learning frameworks for modelling computations in biological neural networks offer an attractive avenue, enabling full observability and controlled interventions. One popular such framework is reservoir computing~\cite{jaeger_2001, maass_2002}, which was originally introduced as a model of spatiotemporal sensory information processing in the brain~\cite{buonomano_maass_2009}. Reminiscent of biological neural networks, the core processing unit of a reservoir computer (RC) constitutes a fixed recurrent neural network (RNN), called the reservoir (Fig.~\ref{fig:introFigure}A). The reservoir receives input from a linear input layer and projects it to a high-dimensional non-linear space, facilitating linear separability~\cite{yan_2024}. Subsequently, the reservoir state is mapped to a desired output via a linear readout layer. Importantly, RC training involves optimising only the readout weights, which allows for fast and efficient training, and the flexibility to use any arbitrary RNN as the reservoir. Notably, there has been a recent trend of using RCs with bio-inspired reservoir topologies, informed by empirical brain connectivity data, to relate structural properties of brain networks to computational functions~\cite{suarez_2021, suarez_2024, damicelli_2022}.

Despite full observability and the possibility to directly assess the contribution of individual artificial neurons to a computation through methods like neuron deactivation (dropout), elucidating how artificial neural networks (ANNs) perform complex computations remains challenging. One reason for this is that relevant information for these computations may not be encoded at the level of single neurons, but at the level of groups of neurons or even the whole network. Such macroscale phenomena that seem irreducible to their constituent parts are characterstic of both artificial and biological neural networks~\cite{luppi_2024_tics}. For instance, in the human brain, information from various sensory modalities is processed in different cortical areas, yet we perceive objects as whole, unified entities rather than disjointed collections of sensory inputs. Exactly how the brain achieves this integration of distributed information streams poses a major question in neuroscience, commonly referred to as the ``binding problem''~\cite{feldman_2013_binding-problem}.

The binding problem exemplifies how the macroscale can possess qualities that are absent at the microscale. This perhaps somewhat paradoxical condition is captured by the concept of emergence~\cite{bedau_2008}, with emergent phenomena being in some sense ``greater than the sum of their parts''~\cite{luppi_2023, barnett_2023}. Emergence offers a promising conceptual approach to addressing the challenges faced by traditional reductionist methods in explaining complex computations in neural networks. However, quantifying the role of emergence in computation has long been limited by the lack of effective measures. 

Recently, Rosas and Mediano et al.~\cite{rosas_mediano_2020} developed a framework that offers a formal definition of causal emergence alongside practical tools for measuring emergence in empirical data. Drawing on the theory of partial information decomposition (PID)~\cite{williams_beer_2010}, the framework demonstrates how macroscale phenomena can have excess causal power over the evolution of the system, beyond what can be accounted for by considering a microscale description of the system (see~\cite{mediano_2022} for a comprehensive review). Please note that in this context ``causal power'' means predictive power, consistent with the Granger interpretation of causality~\cite{bressler_seth_2010}.

PID, which was introduced as an extension of Shannon's information theory, proposes to conceptualise the information that multiple sources provide about a target as a composition of distinct information atoms: i) redundant information that is shared between sources, ii) unique information that is exclusive to each source, and iii) synergistic information that can only be accessed by integrating the information from multiple sources~\cite{williams_beer_2010} (Fig.~\ref{fig:introFigure}B). Building on this approach, Rosas and Mediano et al.~\cite{rosas_mediano_2020} posit that an emergent feature must encode synergistic information about the future of the system.

To illustrate the rationale behind this approach, note that synergistic information only exists at the macroscale -- it is not contained in any subset of system parts -- yet, it is disclosed when all parts are known and considered together. In other words, synergy elegantly accommodates excess causal (i.e., predictive) powers at the macroscale without violating reductionist principles. Hence, the causal emergence framework is entirely compatible with ontological reductionism, while having great potential to provide informative insights into part-whole relationships.

In this study, we explore the hypothesis that complex computations supporting the prediction of environmental variables in biological neuronal networks rely on emergent dynamics. More specifically, we leverage reservoir computing to model the prediction of environmental dynamics in a bio-inspired computational architecture, and we capitalise on the causal emergence framework to examine the relationship between emergence and prediction performance. Notably, we find evidence for a bidirectional coupling between emergence in the system's dynamics and prediction performance. Overall, our results reveal clear empirical benefits of considering emergent dynamics for the study of complex computations in neural networks. In turn, these \textit{in silico} results may provide insights about computation in biological neural networks. 

\section{\label{sec:Results}Results}

\begin{figure*}[t]
    \centering
    \includegraphics{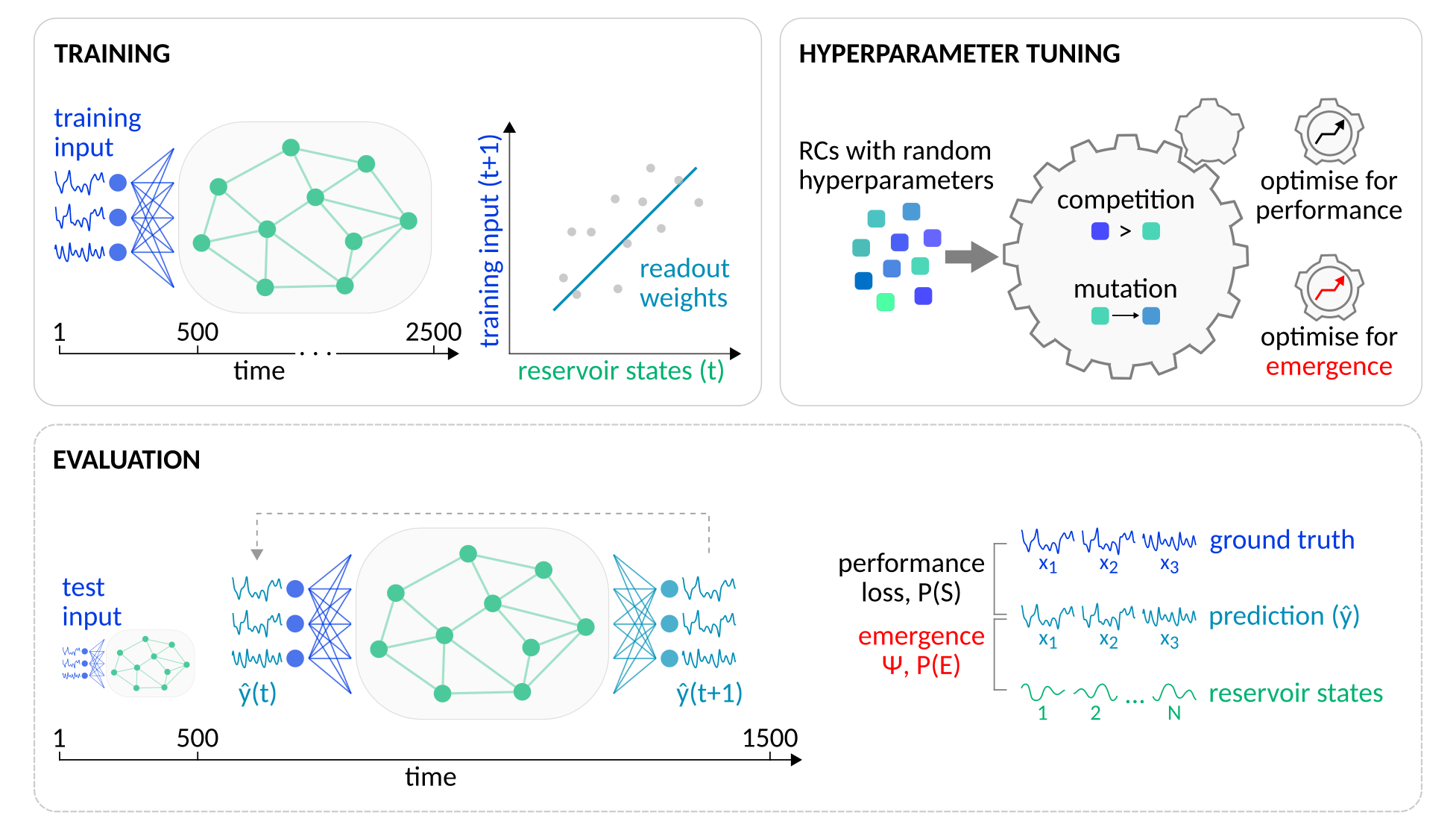}
    \caption{\textbf{Schematic overview of the approach.} Our approach comprises three main operations: training, evaluation, and hyperparameter tuning. Training involves recording the reservoir state trajectory as the RC is fed with some training input, and then computing the readout weights with closed-form linear regression (Tikhonov-regularised ridge regression) to predict the future training input at time t+1 from the current reservoir states at time t. Evaluating the prediction performance and emergent dynamics of a trained RC requires generating predictions by employing an iterative one-step-ahead prediction approach after the RC was fed with the initial 500 time steps of the test input. Prediction performance is assessed by comparing the ground-truth environmental dynamics with the forecast, measuring prediction loss and the probability of a successful prediction $\textrm{P}(S)$ with $S := \textrm{loss}<1$. Emergent dynamics are evaluated by comparing the forecast with the underlying reservoir state trajectory, measuring $\psi$, i.e., the amount of self-predictive information of the forecast that is not contained in individual reservoir neurons. Additionally, we also estimated the probability of emergence $\textrm{P}(E)$ with $E := \psi>0$. Finally, hyperparameter tuning is performed using an evolutionary algorithm to optimise specific RC hyperparameters. We employed different objective functions, either optimising for maximal prediction performance or emergent dynamics.}
    \label{fig:conceptual-overview}
\end{figure*}

To examine the role of emergence in facilitating predictions of environmental dynamics in biological neural networks, we optimised bio-inspired RCs to forecast the trajectories of various chaotic dynamical systems, representing the RCs' environments. Our set of task environments included the well-known Lorenz attractor \cite{lorenz_1963} and five Sprott chaotic flow systems \cite{sprott_1994}. The reservoir connections of our RCs were determined by a human connectome representation describing the anatomical connections between 100 cortical brain regions \cite{schaefer_2018}, which was derived from empirical data of 100 healthy subjects of the Human Connectome Project \cite{vanEssen_2013}.

Our approach for investigating the relationship between prediction performance and emergence in reservoir computing comprised three main operations: training, evaluation, and hyperparameter tuning (Fig.~\ref{fig:conceptual-overview}).

During training, the RC is first ``driven'' with the training input. That is, the reservoir state is iteratively updated as an environmental time series is fed to the network via the linear input weights. Subsequently, the readout weights are analytically solved using closed-form linear regression with Tikhonov regularisation, a.k.a. ridge regression, where the future training inputs at time t+1 are regressed against the generated reservoir states at time t. Unless stated otherwise, our training inputs were 2500 time steps long, and the first 500 time steps were not included in the readout weight fitting.

The evaluation step involves measuring the prediction performance and emergent dynamics of a trained RC. Predictions were initialised by driving the RC with the first 500 time steps of the test input. Then, the actual prediction (forecast) was generated by turning the RC into an autonomous system that drives itself with its own prediction output in an iterative one-step-ahead prediction approach for 1000 time steps. For each evaluation of one RC, multiple (100, unless otherwise specified) forecasts were generated by initialising the RC with different test inputs, and the RC's performance was measured by computing its average prediction loss (see Methods for a precise definition) across all generated forecasts. Additionally, we also derived the probability of a successful prediction $\textrm{P}(S)$ with $S:=\textrm{loss}<1$ as a less outlier-sensitive measure of performance. 

The tendency of an RC to exhibit emergent dynamics was assessed using $\psi$, which provides a lower bound on causal emergence \cite{rosas_mediano_2020} (see Methods). In this study, $\psi$ measures the predictive information of the forecast $\hat{y}(t)$ about its immediate future time point $\hat{y}(t+1)$, minus the total information provided by each individual reservoir neuron considered separately~\footnote{In this study, we only consider first-order $\psi$, which estimates first-order emergence. First-order emergence occurs when the macroscopic feature (i.e. the forecast) predicts itself better than the total information provided by each individual reservoir neuron, considered in isolation. In contrast, full-order emergence requires the forecast to have greater self-predictive power than any possible subset of reservoir neurons.}. Note that $\psi>0$ represents a sufficient, but not necessary, condition for emergence. Negative $\psi$ values can occur if multiple reservoir neurons share redundant information about $\hat{y}(t+1)$, which, however, does not rule out emergence. Thus, we devised an additional metric to assess the tendency of an RC to exhibit emergent dynamics: the probability $\textrm{P}(E)$ of generating a prediction with $E:=\psi>0$. By default, all metrics of performance (loss, $\textrm{P}(S)$) and emergence ($\psi$, $\textrm{P}(E)$) were estimated over 100 test inputs, each being 1000 time steps long (not including the initial 500 time steps of the test input that are fed to the reservoir for initialization). 

Finally, hyperparameter tuning involves using an evolutionary algorithm \cite{harvey_2011} to optimise specific RC hyperparameters, such as topological properties of the reservoir network, that are key for effective training. Mimicking biological evolution, in evolutionary optimization a population of individuals (here RCs) with randomly initialised hyperparameter configurations is evolved towards a specific optimization objective over the course of many generations of competition between individuals and subsequent mutation of the hyperparameter configurations employed by inferior individuals (Fig.~\ref{fig:conceptual-overview}). In the present study, we evolved RCs with two different objective functions: 1) to maximise prediction performance (\textit{optimization objective I}), and 2) to maximise causal emergence (\textit{optimization objective II}).

Overall, this approach allowed us to manipulate performance during hyperparameter tuning and training, and evaluate how these manipulations affect emergence, and vice versa. Specifically, we assessed the effect of hyperparameter optimization for \textit{optimization objective I} and \textit{optimization objective II} on emergent dynamics and prediction performance. We estimated the correlations between performance and emergence across the entire hyperparameter space without optimization. By varying the input sample size during training, we targeted prediction loss and measured the resulting changes in emergent dynamics. Finally, we analysed whether hyperparameter optimization for emergence enhances RC performance in a range of unfamiliar (non-optimised) environments, and whether a human connectome-based reservoir topology promotes emergent dynamics and performance.

\subsection{Prediction performance and emergence}

\begin{figure*}[t]
    \centering
    \includegraphics{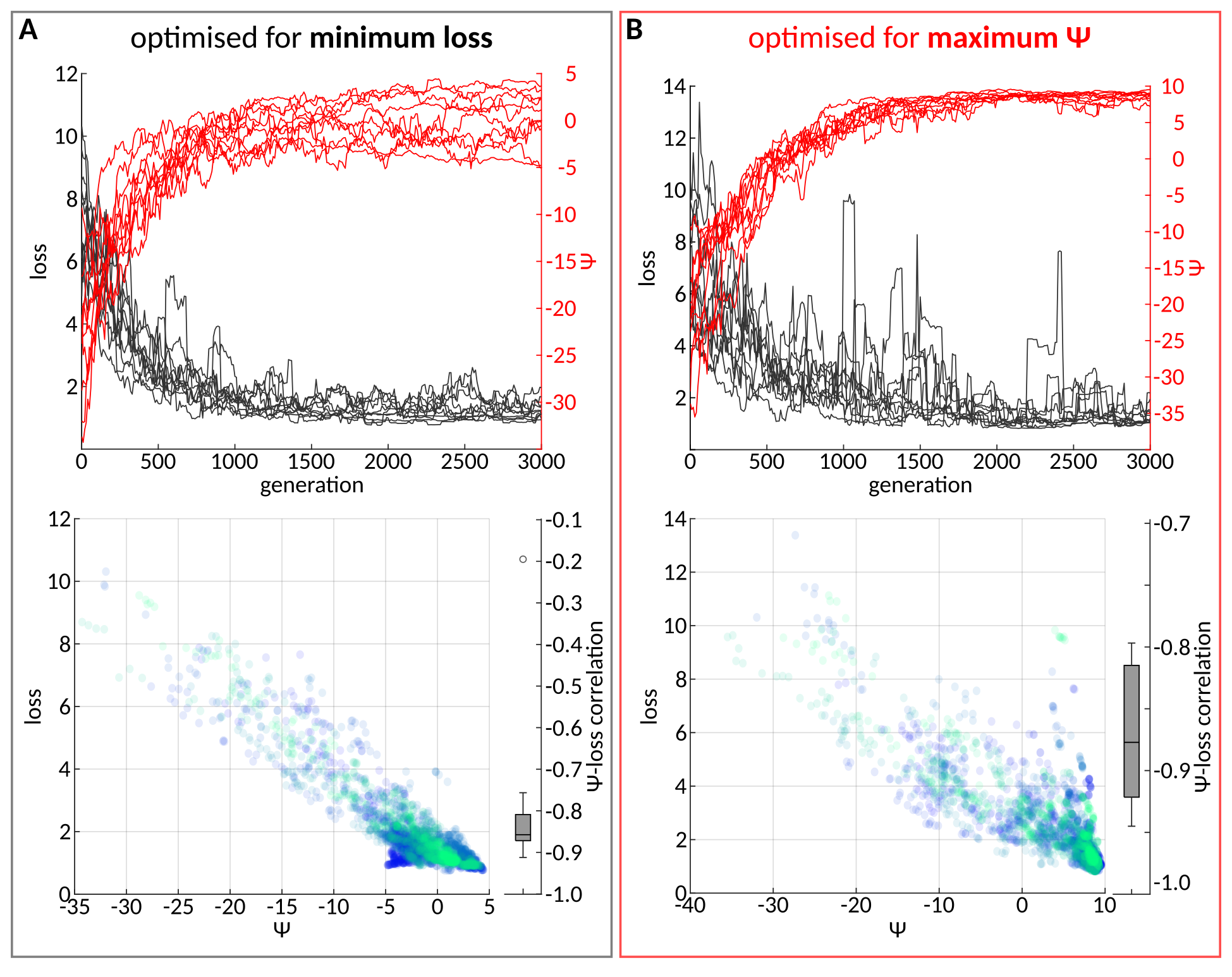}
    \caption{\textbf{Inverse relationship between loss and $\psi$ during evolutionary optimization.} (\textbf{A}) Top: trajectories of average loss (black) and $\psi$ (red) of 10 populations over the course of evolutionary optimization with the objective to minimise loss (\textit{optimization objective I}). Bottom: average loss plotted against average $\psi$ across generations. Each population is plotted in a different colour. The corresponding Spearman correlation coefficients of all populations are summarised in the boxplots (all correlations were significant; $p<0.0001$). (\textbf{B}) Analogous to (A) but for evolutionary optimization with the objective to maximise $\psi$ (\textit{optimization objective II}).}
    \label{fig:analysis01A}
\end{figure*}

\begin{figure*}[t]
    \centering
    \includegraphics{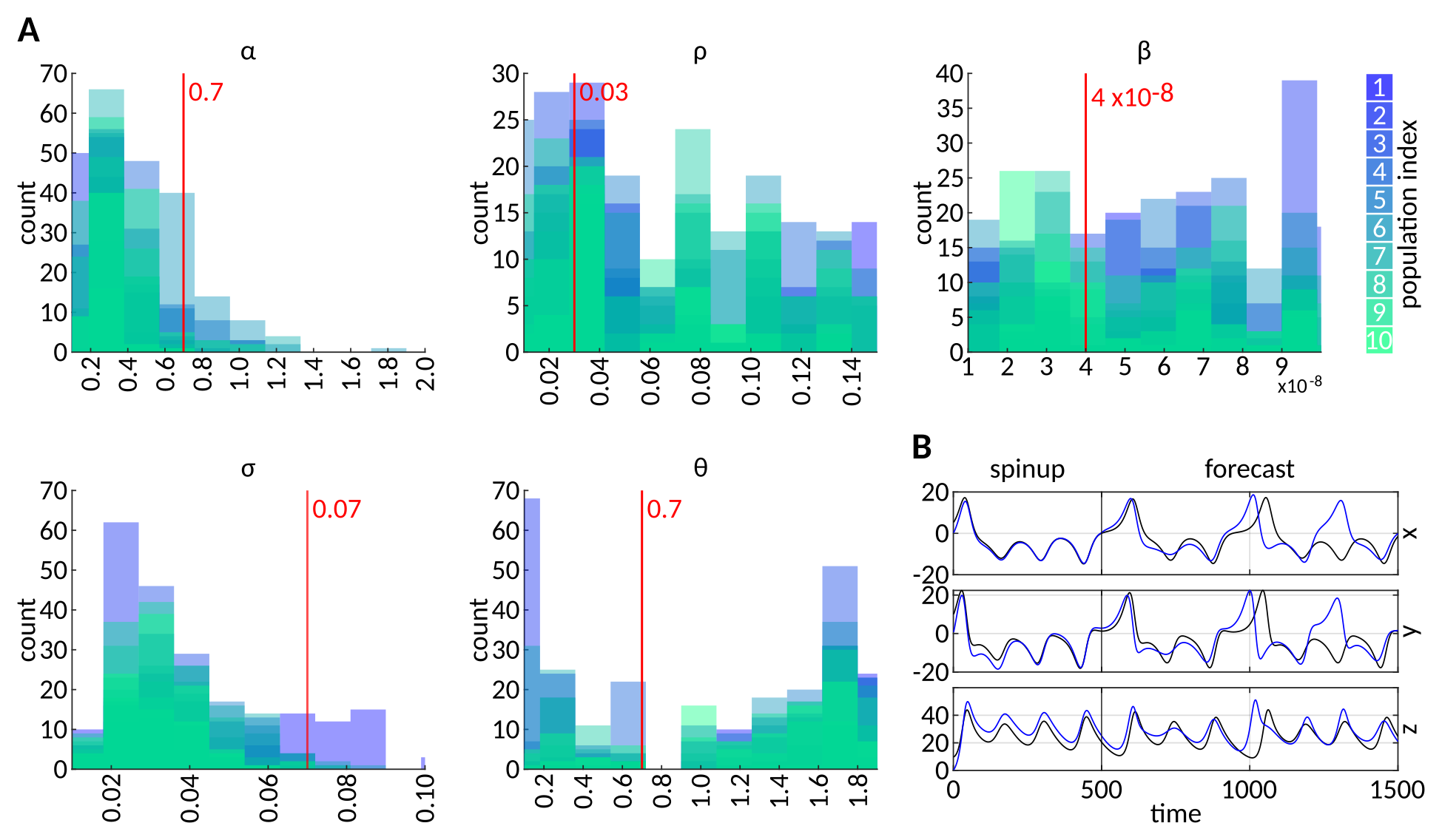}
    \caption{\textbf{Hyperparameter optimization of bio-inspired reservoir computers for minimal loss.} (\textbf{A})  Distributions of hyperparameter values in the final generation of 10 evolved populations. ($\alpha$: the spectral radius of the reservoir adjacency matrix $C$; $\rho$: connection density of $C$; $\beta$: Tikhonov ridge regularisation parameter; $\sigma$: input strength; $\theta$: input bias) The red lines indicate the hyperparameter values of the best solution across all populations. (\textbf{B}) Ground-truth trajectories of each Lorenz variable (black) and RC output (blue), generated by a trained, bio-inspired RC employing the loss-optimal hyperparameters. Black vertical lines demarcate the spinup time from the forecast time. During spinup, the initial 500 time steps of the target environmental trajectory are fed to the reservoir. During forecasting, the RC receives no external input but produces a forecast in a one-step-ahead prediction approach. The readout weights of the RC were computed beforehand on an independently sampled environmental time series for training.}
    \label{fig:S1Fig}
\end{figure*}

\begin{figure*}[t]
    \centering
    \includegraphics{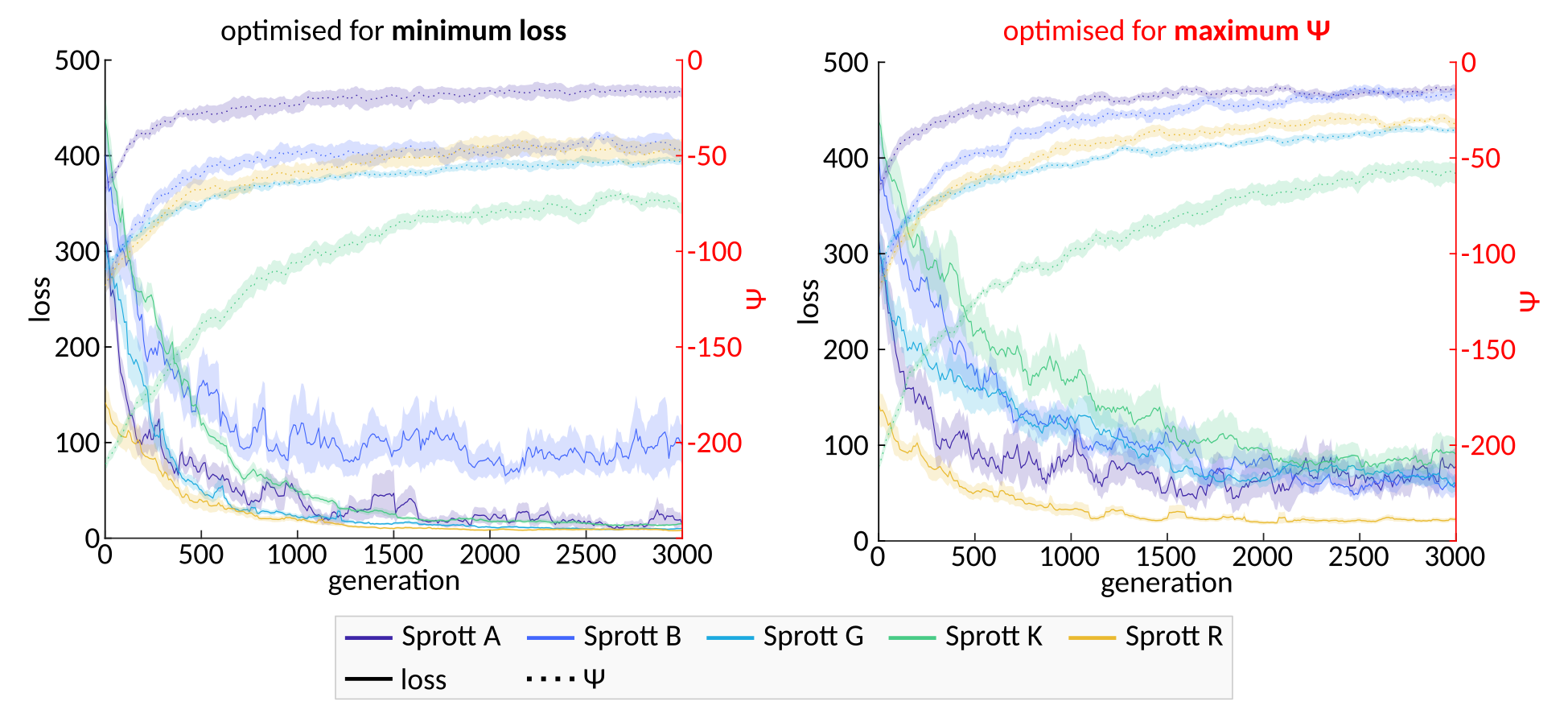}
    \caption{\textbf{Inverse relationship between loss and $\psi$ during evolutionary optimization of bio-inspired reservoir computers.} Left: Mean trajectories of loss (solid line; left y-axis) and $\psi$ (dotted line; right y-axis), averaged across 10 populations per environment (as indicated by the colour and legend), over the course of evolutionary optimization with the objective to minimise loss. Right: Analogous to the right plot but for evolutionary optimization with the objective to maximise $\psi$.}
    \label{fig:S2Fig}
\end{figure*}

To examine the relationship between loss and $\psi$ during hyperparameter tuning, we evolved 10 populations, each comprising 100 RCs, for 3000 generations to the Lorenz task environment, optimising for low loss (\textit{optimization objective I}). As expected, the average population loss decreased over the course of artificial evolution (Fig.~\ref{fig:analysis01A}A), and the best individual across all populations in the final generation was verified to produce a good forecast (Supp. Fig.~\ref{fig:S1Fig}), indicating successful optimization. Notably, the decrease in population loss was consistently paralleled by an increase in population $\psi$ (Fig.~\ref{fig:analysis01A}A). Indeed, we found significant negative correlations ($p<0.0001$; see the correlation coefficients in the bottom panel of Fig.~\ref{fig:analysis01A}A) between the evolutionary trajectories of loss and $\psi$ in all populations. These findings replicated across all tested task environments (Supp. Fig.~\ref{fig:S2Fig}).

A further analysis uncovered that the inverse relationship between loss and $\psi$ is bidirectional: Evolving 10 additional populations, which were initialised identically to those optimised for minimal loss, with the objective to maximise $\psi$ (\textit{optimization objective II}), lead not only to an increase in $\psi$ (as we should expect), but also to a simultaneous decrease in population loss over the course of artificial evolution (Fig.~\ref{fig:analysis01A}B). Notably, optimising for high $\psi$ occasionally resulted in transient spikes in the loss curve, suggesting that $\psi$ maximisation may introduce some instability to the reservoir dynamics, leading to attenuated prediction performance. However, the spikes in loss were short-lived, implying that $\psi$ optimization effectively pruned hyperparameter configurations associated with poor performance.

To ensure that the observed relationship between loss and $\psi$ are not primarily driven by RCs with particularly poor performance and highly synchronised (i.e., redundant) reservoir neurons, we examined the coupling between emergent ($\psi>0$) and successful ($\textrm{loss}<1$) predictions across the hyperparameter search space. To this end, we sampled 4000 RCs with random hyperparameter configurations and estimated their marginal and joint probabilities of producing emergent dynamics and successful predictions for each of the six task environments. Only reservoirs with non-zero probabilities of success and emergence were included in the analysis, leading to $3584\pm153$ (mean$\pm$s.d.) reservoirs per environment.

\begin{figure*}[t]
    \centering
    \includegraphics{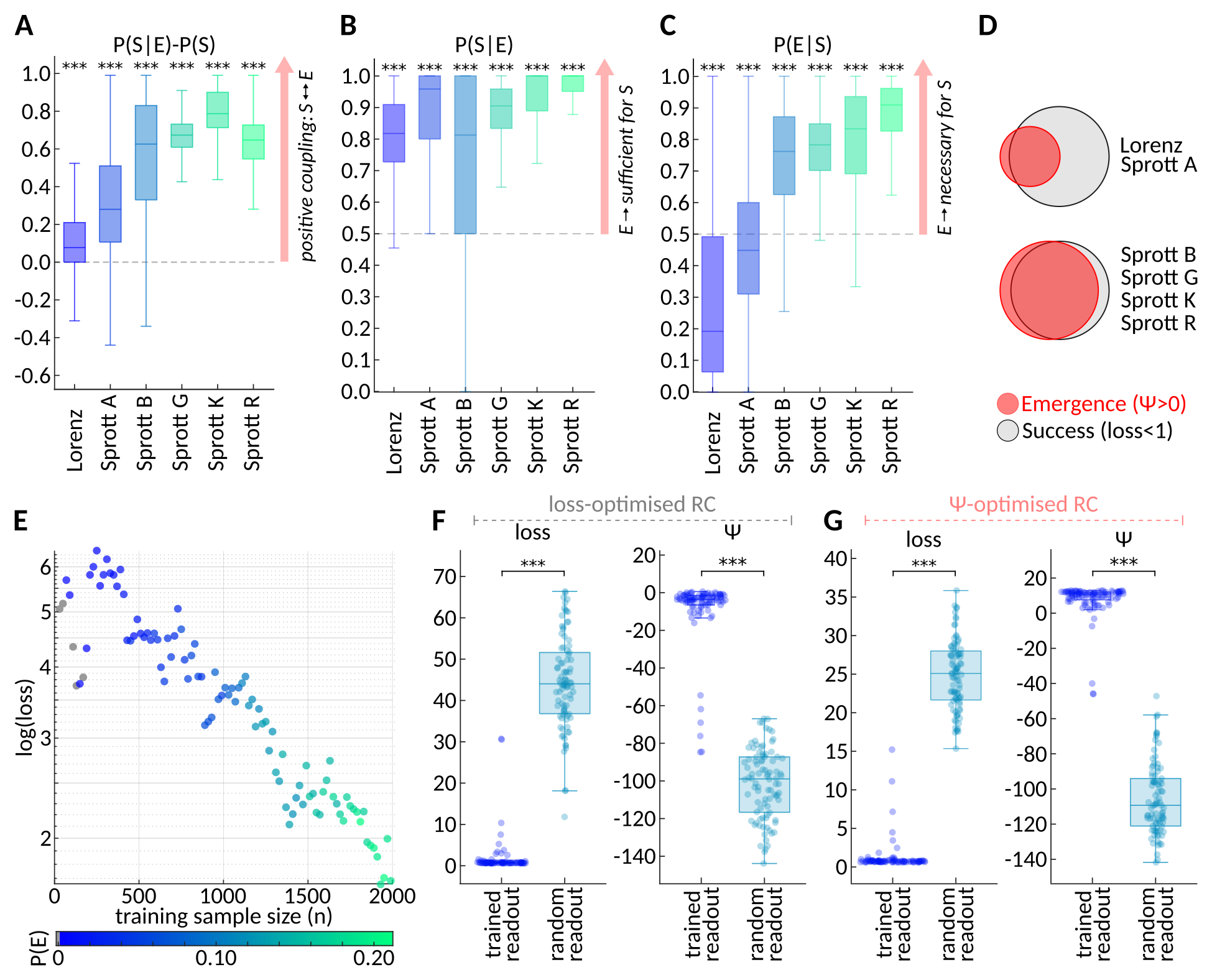}
    \caption{\textbf{Relationship between emergence and prediction performance across hyperparameter configurations and training sample sizes.} (\textbf{A}) 4000 RCs with randomly sampled hyperparameter configurations were evaluated 100 times on each of the six task environments. The boxplots summarise for each environment the difference in the probability of a successful prediction (loss$<$1) when conditioning on emergence ($\psi>0$). RCs with $\textrm{P}(S)=0$ or $\textrm{P}(E)=0$ were excluded from the analysis, resulting in $3584\pm153$ (mean$\pm$s.d.) data points in each boxplot. (\textbf{B}) Fraction of successful predictions over emergent predictions, and (\textbf{C}) fraction of emergent predictions over successful predictions for each task environment, derived from the same data as (A). (\textbf{D}) Schematic Venn diagrams of emergence and prediction success in the different task environments, inferred from the results in (B-C). (\textbf{E}) Average log(loss) and probability of emergence achieved by the loss-optimised RC when trained with different training sample sizes ($n$). (\textbf{F}) Comparison of loss and $\psi$ values of predictions produced by the loss-optimised RC employing trained versus randomised readout weights. (\textbf{G}) Analogous to (F) but the predictions were produced by the $\psi$-optimised RC. Asterisks (***) denote statistically significant differences after FDR-correction ($p<0.0001$).}
    \label{fig:analysis01BCD}
\end{figure*}

Remarkably, emergent predictions were found to be significantly more likely to be successful ($\textrm{P}(S|E)-\textrm{P}(S)>0$) (Fig.~\ref{fig:analysis01BCD}A). Furthermore, for all task environments the fraction of successful predictions over emergent predictions ($\textrm{P}(S|E)$) was significantly greater than chance (0.5), with the median probabilities consistently exceeding 0.8 (Fig.~\ref{fig:analysis01BCD}B), suggesting that emergence ($\psi>0$) represents an almost sufficient criterion for a prediction to be successful in all tested environments (Fig.~\ref{fig:analysis01BCD}D). Moreover, for the environments Sprott B, G, K and R, we found that the fraction of emergent dynamics over successful predictions $\textrm{P}(E|S)$ was also significantly and substantially greater than 0.5 (Fig.~\ref{fig:analysis01BCD}C), indicating that for forecasting the dynamics of these environments, emergent dynamics are not only near sufficient but also almost necessary for prediction success (Fig.~\ref{fig:analysis01BCD}D). Given that $\psi>0$ represents a sufficient criterion for emergence, these results complement our previous findings with the supporting evidence that prediction performance is indeed linked to emergence, and not solely to reduced correlation of reservoir neuronal activity leading to higher but possibly still negative $\psi$ values.

Our previous analyses focused on the relationship between emergence and prediction performance across different hyperparameter configurations. Next, we aimed to explore the relationship between emergence and prediction performance within a single RC, while varying the size of the training dataset. Specifically, we trained an RC employing the hyperparameter configuration that yielded the lowest loss across all 10 loss-optimised populations. The RC was trained on 100 different training inputs from the Lorenz environment, and each training was evaluated on the same set of 10 test time series. Crucially, instead of training the RC on the entire length of the input data (2000 time steps), we trained it on truncated versions with varying sample sizes, i.e. numbers of time steps ($n = \{10, 20, ..., 2000\}$).

Figure~\ref{fig:analysis01BCD}E displays the average log(loss) and $\textrm{P}(E)$ across all evaluations for each training sample size $n$. As expected, loss decreases with increasing training sample size. Interestingly, larger training sample sizes not only led to lower loss, but also to a higher probability of emergence, suggesting that fitting the output weights with more data results in enhanced readout of emergent dynamics.

A further analysis revealed that those emergent dynamics are indeed likely to encode task-relevant information (Fig.~\ref{fig:analysis01BCD}F): We trained and subsequently evaluated the same loss-optimal RC 100 times, each time on a new set of 1 training and 1 test time series from the Lorenz environment. Additionally, during each repetition we also computed the loss and $\psi$ with respect to a random readout of the RC, which was obtained by randomly permuting the trained readout weights, while preserving the weight distributions of the mappings from reservoir neurons to each of the target variables. Unsurprisingly, random readouts yielded significantly greater loss than the predictions generated with trained readout weights ($g=-1.70$, $t=-12.33$, $p<0.0001$, df=99). More interestingly, random readouts were also associated with significantly and substantially lower $\psi$ values ($g=3.47$, $t=37.63$, $p<0.0001$, df=99). Of note, this trend was found to be even more pronounced when repeating the analysis with an RC that employed $\psi$-optimised, rather than loss-optimised, hyperparameters, where trained readouts nearly always achieved $\psi>0$, whereas random readouts never did (Fig.~\ref{fig:analysis01BCD}G) (loss comparison: $g=-1.00$, $t=-7.36$, $p<0.0001$, df=99; $\psi$ comparison: $g=5.96$, $t=51.37$, $p<0.0001$, df=99).

\subsection{Transfer learning to unfamiliar environments}

\begin{figure*}[t]
    \centering
    \includegraphics{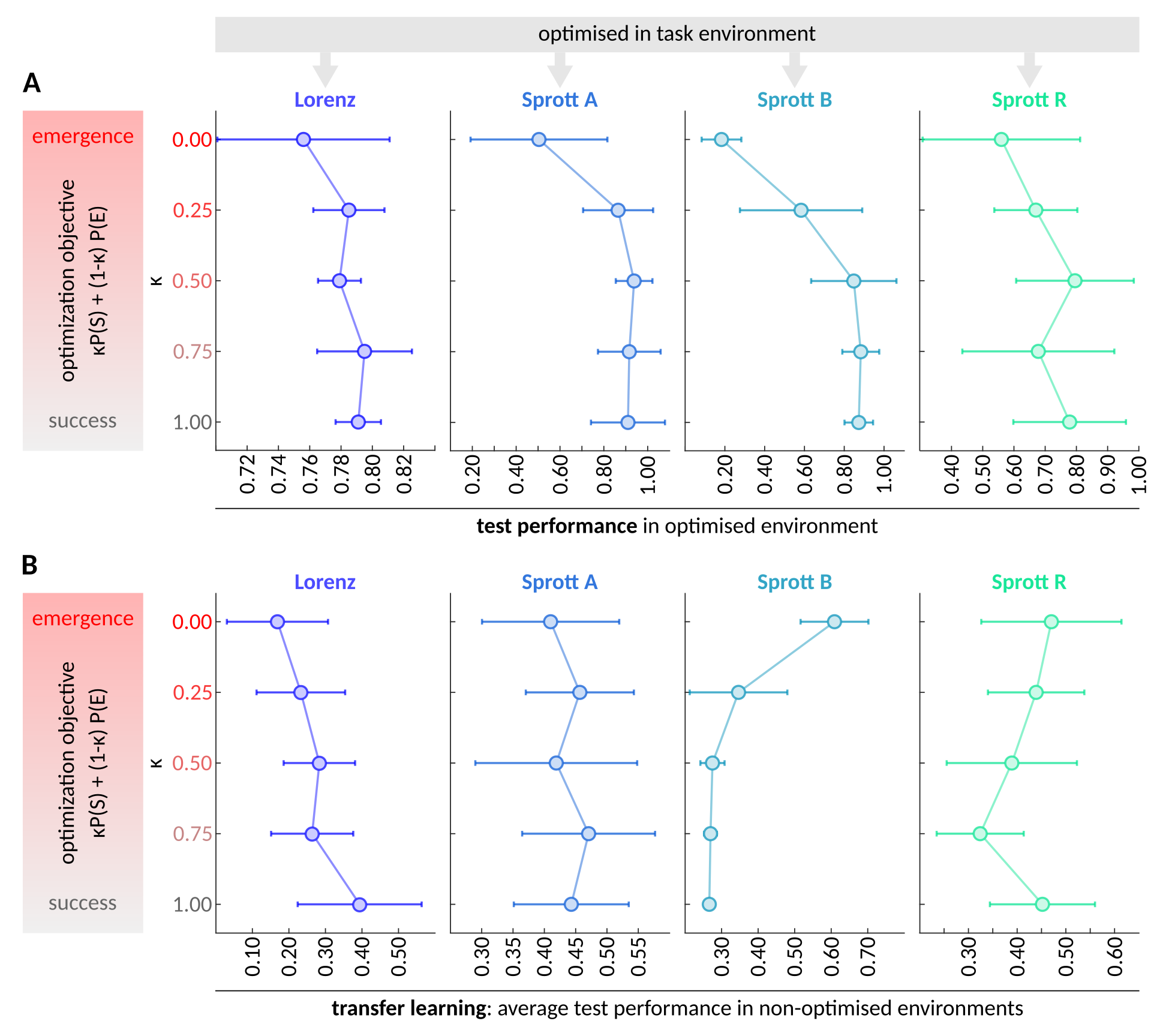}
    \caption{\textbf{Test performance of reservoir computers, evolved to maximise performance or emergence, across optimised and non-optimised environments.} (\textbf{A}) Means $\pm$ standard deviations of the test performances ($\textrm{P}(S)$) of the fittest RCs that were evolved to maximise $\kappa \textrm{P}(S)+(1-\kappa) \textrm{P}(E)$ in a particular environment, indexed by the column and the line plot colour, and evaluated on the same environment. Each dot in the line plot summarises the performances of the fittest RC from each of the 10 populations that were evolved for the given environment (column) and $\kappa$ value (y-axis). (\textbf{B}) Analogous to (A) but the x-axis displays the average test performance of the RCs across all other (non-optimised) environments.}
    \label{fig:analysis02}
\end{figure*}

We next investigated whether hyperparameter optimization to maximise emergence may offer advantages over performance-focused optimization in the challenging context of transfer learning. Specifically, based on previous work showing that synergistic information processing facilitates general-purpose learning in artificial neural networks~\cite{proca_2022}, we wondered whether hyperparameter tuning towards a higher tendency to exhibit emergence -- which, by definition, relies on synergistic information~\cite{rosas_mediano_2020} -- would prime RCs to perform well on a range of unfamiliar environments (that they were not evolved to).

To test this hypothesis, we evolved RCs to maximise various weighted combinations of $\textrm{P}(S)$ and $\textrm{P}(E)$. More precisely, our optimization objective was defined as $\kappa*\textrm{P}(S)+(1-\kappa)*\textrm{P}(E)$, where $\kappa$ ranged from 0.0 to 1.0 in increments of 0.25. Hence, $\kappa=0.0$ corresponds to 100\% optimization for $\textrm{P}(E)$ and 0\% for $\textrm{P}(S)$, and vice versa in the case of $\kappa=1.0$. For each $\kappa$ value, we evolved 10 populations for 3000 generations, each comprising 100 randomly initialised RCs, to a given task environment. The analysis was conducted with four different task environments, the Lorenz attractor and selected Sprott systems (A, B, R), which were chosen based on the effectiveness of the hyperparameter optimization for these environments in the previous analysis. Finally, the best hyperparameter configuration from each population, which was evolved with a specific $\kappa$ value and to a specific environment $i$, was once evaluated on a new set of 100 test time series from the optimised environment $i$, and additionally on a set of 100 test time series from each of the 3 other environments $j \neq i$.

Figure~\ref{fig:analysis02} summarises the test performance ($\textrm{P}(S)$) in the optimised environment (top row) and the average test performance across all non-optimised environments (bottom row) of the fittest RC from each evolved population. As expected, we found that optimising for $\textrm{P}(S)$ alone (i.e. $\kappa$=1.0), tends to yield the best test performance in the optimised task environment. When the RC is evaluated on a \textit{different} task environment, optimising for $\textrm{P}(E)$ may be advantageous in some cases - although not all. While further investigation is needed, the fact that optimization for emergence alone yielded solutions with competitive, and occasionally superior, test performance in non-optimised environments hints at the potential value for future research into the conditions where emergence optimization may be beneficial.

\subsection{Bio-inspired versus random reservoir topologies}

In standard reservoir computing, reservoir neurons are typically randomly connected such that every neuron is connected to all other neurons with a fixed probability $p$ \cite{jaeger_2001, maass_2002}. Recently, there has been a trend towards leveraging RCs with bio-inspired reservoir topologies, where the reservoir neurons are connected like brain regions in empirically derived structural brain networks \cite{suarez_2021, damicelli_2022, suarez_2024}. This approach has been proposed for relating structural properties of brain networks to computational functions.

The RCs analysed above were all constructed using empirical brain network data. To test whether this mattered, we compared the prediction performance and emergence of randomly connected RCs with that of our bio-inspired RCs. More precisely, for each reservoir type (human connectome or random network) and for each of the six task environments, we evolved 10 populations of 100 RCs for 3000 generations, optimising for minimal prediction loss. To improve comparability, we ensured that for each population of bio-inspired RCs in a given environment, there was a sister population of randomly connected RCs in the same environment, which was identically initialised at generation 0.

\begin{figure*}[t]
    \centering
    \includegraphics{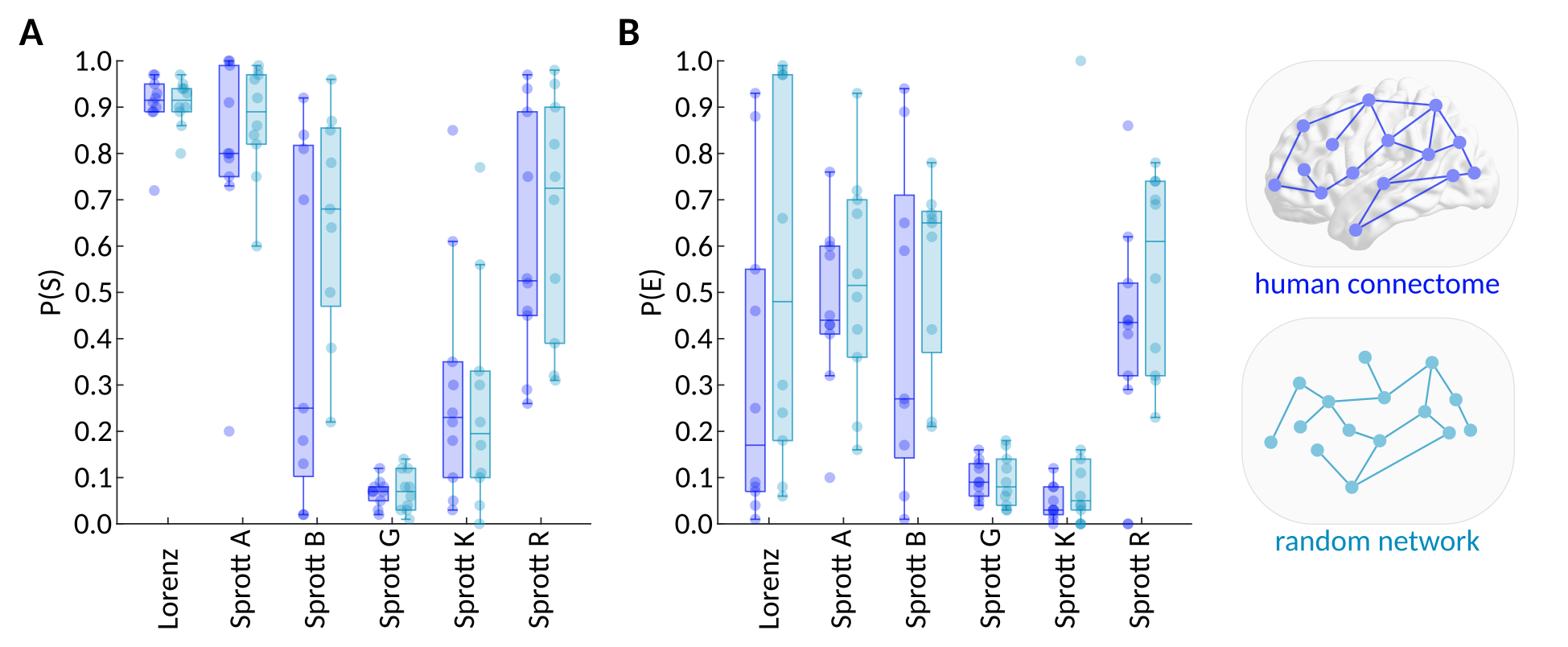}
    \caption{\textbf{Comparing prediction performance and emergent dynamics of bio-inspired and randomly connected reservoir computers.} The analysis involved evolving 10 populations for each task environment and reservoir network type. (\textbf{A}) Performance ($\textrm{P}(S)$) of the fittest RC from each evolved population of RCs with bio-inspired reservoir topology, informed by the human connectome, (dark blue) and RCs with randomly connected reservoirs (light blue). (\textbf{B}) Emergence ($\textrm{P}(E)$) of the same RCs as in (A). There were no significant differences in performance or emergence between bio-inspired and randomly connected RCs.}
    \label{fig:analysis03}
\end{figure*}

Evaluating the best solution from each evolved population on 100 test time series from the optimised environments revealed no significant differences between bio-inspired versus randomly connected RCs in success or emergence probability (Fig.~\ref{fig:analysis03}). 

Previous studies comparing RCs with bio-inspired, human connectome-based reservoirs to those with randomly connected reservoirs have yielded seemingly conflicting results: Some found worse performance for bio-inspired RCs, or equal performance when a slight weight randomization is employed~\cite{damicelli_2022}, while others reported better performance for bio-inspired RCs~\cite{suarez_2021}. In this context, our results emphasise the need for further research to determine when reservoir computing represents a suitable model for computation in the human brain. Which aspects are captured, and which ones are not? Furthermore, the mentioned studies~\cite{damicelli_2022, suarez_2021} and our study, all differ in their methods, underscoring the importance of identifying which approaches are appropriate for specific types of questions.

\begin{figure*}[t]
    \centering
    \includegraphics{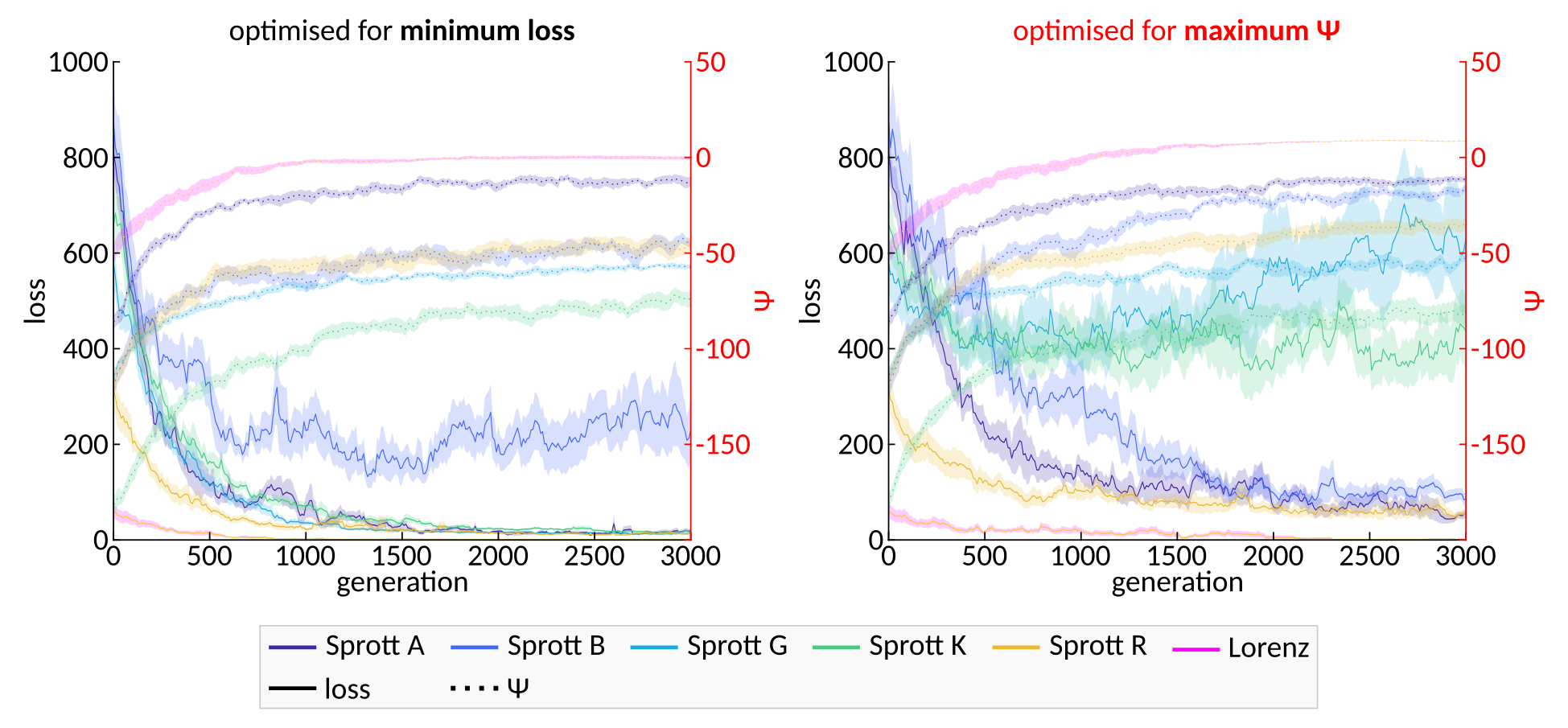}
    \caption{\textbf{Inverse relationship between loss and $\psi$ during evolutionary optimization of randomly connected reservoir computers.} Left: Mean trajectories of loss (solid line; left y-axis) and $\psi$ (dotted line; right y-axis), averaged across 10 populations per environment (as indicated by the colour and legend), over the course of evolutionary optimization with the objective to minimise loss. Right: Analogous to the right plot but for evolutionary optimization with the objective to maximise $\psi$.}
    \label{fig:S3Fig}
\end{figure*}

Besides pointing out the crucial need for those methodological clarifications, our result suggests that the relationship between emergence and prediction performance observed in previous analyses is not specific to the bio-inspired RC architecture employed in our study, but rather it may represent a more general principle. Indeed, we confirm that the strong inverse correlations between $\psi$ and loss during hyperparameter optimization are recapitulated in randomly connected RCs (Supp. Fig.~\ref{fig:S3Fig}).

Finally, it is noteworthy that, compared to other environments, the optimization in the Lorenz and Sprott A environments yielded markedly more successful RCs (Fig.~\ref{fig:analysis03}A). These two environments were also previously found to stand out as the only environments for which emergent dynamics tended to be not necessary, albeit almost sufficient, for prediction success (Fig.~\ref{fig:analysis01BCD}C). This hints at a potential link between the reliance of prediction success on emergent dynamics and the complexity of the prediction task at hand, which may be explored by future work.

\section{Discussion}

Our study reveals a bidirectional relationship between prediction performance and emergence in reservoir computing, which is consistent across task environments and reservoir topologies. Optimising hyperparameters for performance enhances emergent dynamics, and vice versa. Emergent dynamics are almost sufficient for prediction success in all environments and nearly necessary in most. Training RCs with larger sample sizes increases the readout of emergent dynamics, indicating that task-relevant information is encoded in synergistic neuronal interactions. Additionally, our results suggest that emergence-based hyperparameter optimization in one environment may promote performance in other, non-optimized environments.

\subsection{Weak learners collectively become strong through emergence}

Our findings highlight the importance of considering emergent dynamics when studying computation in weak learner systems like RCs. But what exactly do we gain from an emergence-based approach? Emergence-based approaches typically provide tools for identifying optimal representations, or ``coarse-grainings,'' of the system that enable better prediction and/or control of its behaviour -- with different definitions of emergence offering complementary insights. For instance, according to Barnett et al.'s dynamical independence framework~\cite{barnett_2023} emergence occurs when a coarse-graining is optimally predicted based on its own history alone, without requiring further knowledge about the underlying microscale dynamics. Furthermore, Hoel et al. introduced the notion of causal (i.e., interventionist) emergence~\cite{hoel_2013, hoel_2016}, where the system's response to interventions can be predicted more easily in the macroscale than in the microscale. This approach reveals coarse-grainings that allow for more effective control over the system's behaviour. 

In our study, we used the causal emergence framework by Rosas and Mediano et al.~\cite{rosas_mediano_2020}, where the macroscale representation has excess self-causing powers due to synergistic interactions at the microscale. Unlike Hoel et al.'s framework~\cite{hoel_2013, hoel_2016}, which is based on interventionist causality~\cite{pearl_2018}, the notion of causal emergence in our study is grounded in Granger causality~\cite{bressler_seth_2010}. Hence, causality ought to be understood in the predictive sense, i.e., the macroscale possesses self-predictive powers. Yet, unlike the dynamical independence framework~\cite{barnett_2023}, Rosas and Mediano et al.'s approach~\cite{rosas_mediano_2020} requires emergent features to have self-predictive powers that \textit{exceed} what is encoded in the microscale dynamics. This makes it particularly suitable for studying time series prediction in neural networks because it allows for the detection of higher-order interactions between neurons that carry task-relevant information.

To understand why synergy-based causal emergence may be key to enabling environmental predictions in reservoir computing, it is useful to conceptualise each reservoir neuron as a weak learner with limited predictive information about the environment. In the worst-case scenario, all neurons are perfectly synchronised (fully redundant), rendering the predictive power of the reservoir reducible to that of a single neuron. If instead some neurons provide unique information about the environment, the reservoir can achieve improved predictive power by aggregating the redundant and unique information from each neuron. Additionally, task-relevant information may also be encoded synergistically in any possible combination of neurons. Thus, exploiting synergistic information dynamics allows for greatly amplified representational power, effectively uniting weak learners to form a powerful prediction machine with greater computational capacity than the sum of its parts.

Based on these considerations, one might expect RCs to rely more heavily on emergent dynamics for more challenging tasks. Our findings are in line with this idea. Some other work is also supportive: one study, evolving ANNs to task environments of varying complexity, demonstrated that integrated information, a metric related to synergy~\cite{mediano_2021_taxonomy, luppi_2021_bit}, increases over the course of artificial evolution, with the most significant increases occurring in more complex environments~\cite{albantakis_2014} (see also \cite{seth_2005_causal_connectivity} for an earlier example, showing increased (Granger) causal connectivity in networks evolved in complex relative to simple environments). One speculation arising from these data is that reliance on synergistic information in neural networks may follow an inverted-U shape as task difficulty increases, peaking at the point where the network reaches its computational capacity. Additional research is needed to test this hypothesis.

Our results may seem to suggest that maximising emergence, and thereby maximising synergy over redundancy, always leads to improved performance. However, previous studies indicate a trade-off between synergy and redundancy regarding prediction performance~\cite{proca_2022, varley_2024, luppi_2024_tics}. Greater synergy enhances the capacity to integrate information but can lead to unstable, chaotic dynamics~\cite{proca_2022, varley_2024}. Conversely, greater redundancy provides stable dynamics and robustness to perturbations but results in poor information integration~\cite{proca_2022, varley_2024, luppi_2024_tics}. This suggests that the relationship between emergence and performance may in fact be non-linear. Indeed, although maximising $\psi$ reduces overall loss, we observed transient spikes in the evolutionary loss trajectories, suggesting that certain hyperparameter configurations yield poor performance despite emergent dynamics. Notably, these spikes were especially prominent when $\psi$ exceeded the levels achieved by RCs optimised for performance (low loss) rather than emergence (high $\psi$). Overall, we conclude that optimal performance requires i) a balance between synergy and redundancy, and ii) significantly more synergistic dynamics than randomly initialised RCs.

\subsection{Possible implications for biological neural networks}

Our findings allow for several speculations about the relationship between emergence and prediction performance in biological neural networks. 

First, we found that learning a mapping from neural activity to environmental dynamics leads to an increased readout of task-relevant synergistic information, and that emergent dynamics may be especially important for solving challenging tasks. Given these results, we should expect to find emergent dynamics particularly in brain areas that are implicated in learning and performing complex tasks such as the higher-level association cortices. In support of this idea, recent work revealed a gradient of increasing synergistic compared to redundant dynamics from lower-level sensory to higher-level association cortices~\cite{luppi_2022_synergistic-core, varley_2023}. Future studies could investigate the levels of synergistic or emergent dynamics in empirical neuroimaging data from subjects performing cognitive tasks with varying degrees of difficulty.

Second, the robust correlation between emergence and prediction performance during evolutionary hyperparameter optimization raises the  possibility that human brain evolution may have been influenced by emergence-favouring selection pressure. Notably, a recent study showed that the proportion of synergistic interactions between brain regions is significantly higher in humans compared to macaques, while the proportion of redundant interactions remains preserved~\cite{luppi_2022_synergistic-core}. The same study also found that brain regions with highly synergistic dynamics have undergone the greatest expansion during human evolution and display the highest synaptic density and plasticity. Furthermore, dendrites of human pyramidal neurons are capable of purely synergistic computations such as separating linearly non-separable inputs - a feature that has not been shown in non-human neurons~\cite{gidon_2020, rosas_2020_operational, luppi_2024_tics}. 

Interestingly, a recent theoretical account proposed a potential mechanism for these empirical observations: greater synaptic turnover, and ultimately higher synergy, may have been facilitated by metabolites downstream of a non-oxidative glycolysis pathway that brain regions increasingly relied on as their evolutionary expansion outpaced the oxygen delivery capacity of the vascular system~\cite{luppi_2024_oxygen}. Given these findings, it would be interesting to compare the synergistic or emergent dynamics in RCs with reservoir topologies of different species, perhaps following the approaches of Damicelli et al.~\cite{damicelli_2022}, or using the toolbox of Suárez et al.~\cite{suarez_2024}.

If the human brain's structure was indeed shaped by evolutionary pressure to support emergent dynamics, what are the precise structural properties that promote these dynamics? The architecture of the human brain exhibits a number of features, including modular organisation and small-world character, that clearly distinguishes it from random networks~\cite{bullmore_sprons_2009}. Yet, our results did not show a significant difference in emergence or performance between RCs employing the human connectome versus Erdös-Réyni type random networks as reservoirs. However, Suárez et al., comparing different randomised versions of the human connectome, did find that preserving the modular organisation and topology of the human connectome was essential for optimal RC performance on a memory task~\cite{suarez_2021}. Additionally, two independent whole-brain modelling studies demonstrated that changes in emergent or synergistic dynamics in healthy ageing volunteers~\cite{gatica_2022} and coma patients~\cite{luppi_2023} could be explained by underlying structural changes in their connectomes. These findings suggest that the human brain's structure promotes both computational performance and emergent dynamics, which our method may have failed to pick up on. Replicating Suárez et al.'s study~\cite{suarez_2021} while measuring both performance and emergence could provide further insights.

Finally, our analysis provokes the question: was the development of emergence-promoting structural properties in biological brains accelerated by evolutionary pressure to maximise predictive power over diverse environments? While our evidence for emergence enhancing performance in unfamiliar environments is preliminary, it aligns with the findings from Proca et al., who trained ANNs on a diverse set of tasks, demonstrating the critical role of synergy in facilitating multi-purpose learning~\cite{proca_2022}. Exploring the conditions under which optimising for emergence enhances transfer learning abilities to unfamiliar task environments promises to yield important insights not only into brain evolution, but also for developing improved artificial intelligence.

\subsection{Limitations and future work}

While computational models offer full observability (in contrast to the drastic partial observability of \textit{in vivo} biological networks), their major drawback lies in their inherent abstraction from the real-world system of interest. Our finding that bio-inspired reservoir computers with human connectome topology fail to outperform their randomly connected counterparts implies that critical aspects of brain computation are not captured by our model -- at least not at the resolution of macroscale brain networks where reservoir neurons represent brain regions. Notably, reservoir computing theory indicates that RCs rely on rich internal representations, which partly depend on some randomness in the reservoir’s connections~\cite{farrell_2023}. This may explain why the specific topology of the human connectome did not improve prediction performance in our study. However, we do note that others have arrived at opposite conclusions~\cite{suarez_2021}, highlighting the need for further research to resolve these conflicting results. In the meantime, replicating our analysis in an alternative bio-inspired computational framework, such as spiking neural networks~\cite{tan_2020_snn}, could fruitfully complement our findings. Additionally, invasive neuroimaging techniques like optogenetics~\cite{joshi_2020_optogenetics} and 2-photon calcium imaging~\cite{grienberger_2022_ca-imaging} can be employed to enhance the observability of biological neural networks and measure emergent dynamics during task performance in animal studies.

Another limitation of our study is that it does not prove a causal relationship between emergence and prediction performance. The interventionist approach to causality, widely considered the gold standard, establishes causation if a controlled intervention on X elicits a response in Y~\cite{pearl_2018}. While our computational model theoretically allowed for controlled interventions, emergence is a property of the system’s dynamics, which depend on various factors, rendering a specific perturbation of emergence infeasible. This raises the possibility that we optimised a hidden variable, affecting both performance and emergence without a direct causal link between the two. 

However, two arguments support a causal relationship. Firstly, emergence and performance were bidirectionally coupled not only during evolutionary hyperparameter optimization but also when sampling across the entire hyperparameter space, where emergence was nearly sufficient and necessary for prediction success in most environments. Secondly, randomly permuting the readout weights of a trained RC, which dramatically impairs prediction performance, resulted in highly non-emergent dynamics. This suggests that the observed emergence is not merely due to autocorrelation effects in the RC’s forecast but encodes task-relevant information. In essence, when we detect emergence in our study, it means that the RC utilises synergistic information to compute the next prediction output. We show that this coincides with good performance and that the synergistic information is likely task-relevant. Note however, that these results may depend on our selection of tasks. Future analytical work should identify the specific conditions that make emergence necessary for prediction success to provide a more mechanistic understanding and reveal potential causal relationships.

The lack of consensus on defining the partition of information into redundant, unique, and synergistic atoms poses a limitation on all studies using PID~\cite{gutknecht_2023, kay_2022}. However, although causal emergence in our study is defined based on PID~\cite{rosas_mediano_2020}, the metric we used to measure emergence, $\psi$, does not require the computation of PID information atoms. Instead, it relies on the well-established Shannon mutual information~\cite{rosas_mediano_2020}. 

On the downside however, $\psi$ only provides a lower bound on emergence. That is, while $\psi>0$ is sufficient for emergence, it is not necessary. This may have masked part of the effect in our analysis, suggesting that the true relationship between prediction success and emergent dynamics could be even stronger than measured. Future research should aim to develop more exact measures of emergence that are equally efficient to compute.

Furthermore, $\psi$ does not provide insights into the precise causal architecture underlying the observed emergent dynamics. Tantalisingly, recent work has introduced a framework, utilising $\epsilon$-machines to map emergent mechanisms across multiple scales~\cite{rosas_2024_software}. Although estimating $\epsilon$-machines from empirical data remains challenging in practice, a promising approach employing kernel methods has been proposed~\cite{brodu_crutchfield_2021}. Applying this method to our reservoir computing paradigm could elucidate emergent mechanisms during environmental prediction.

\subsection{Conclusion}

We have uncovered a (causal) bidirectional relationship between causal emergence and prediction performance in a bio-inspired reservoir computing model of forecasting environmental dynamics. This emphasises the need for leveraging tools capable of capturing synergistic higher-order interactions, such as the ones used here, when studying computation in biological neural networks. Future research expanding on our work should focus on identifying characteristic features of biological neural networks that promote emergent dynamics. Additionally, understanding the conditions under which optimising for emergence enhances transfer learning to unfamiliar environments remains an important open question. Addressing these questions could not only provide insights into brain evolution, but also holds potential for advancing artificial intelligence.

\section{Methods}

\subsection{Causal emergence}
Emergence was measured using the framework of \cite{rosas_mediano_2020}. The framework posits that a macro-scale feature $V_{t}$ is an emergent feature of a multivariate system $X_{t}$ if i) it is supervenient on $X_{t}$, and ii) it predicts the future of $X_{t}$ better than any subset of system parts. The condition of supervenience requires $V_{t}$ to be fully determined by $X_{t}$ at any time point $t$, such that there exists a possibly noisy function $F(X_{t})=V_{t}+\epsilon$, where $\epsilon$ is an independent noise term.

More formally, this notion of causal emergence can be expressed in the language of PID~\cite{williams_beer_2010}: by decomposing the information that the sources $V_{t}$ and all subsets $X_{t}^{\alpha}$ of the $N$-dimensional system $X_{t}$ with $\alpha \subseteq A$ and $A = \{1, 2, \ldots, N\}$ at time $t$ provide about the target $X_{t'}$ with $t' > t$, we obtain redundant, unique and synergistic information. Causal emergence can then be quantified as the unique information of the source $V_{t}$ about $X_{t}$ that is not contained in any $X_{t}^{\alpha}$ for all $\alpha$. If this quantity is greater than zero, then $V_{t}$ is said to be emergent. $V_{t}$ is emergent iff:

$$\textrm{Un}(V_{t}; X_{t} | X_{t}^{\alpha})>0 \textrm{ } \forall \textrm{ } \alpha $$

Given the super-exponential growth of the number of possible subsets subsets $X_{t}^{\alpha}$ with increasing system size, Rosas and Mediano et al.~\cite{rosas_mediano_2020} introduced the notion of $k$-th order causal emergence, which occurs when $V_{t}$ encodes information about the future of $X_{t}$ beyond what is contained in any set of $k$ or less parts of the system with $k<N$. Note that it is generally more difficult for a feature to meet the requirements for causal emergence of higher orders. In this work, we restrict our analysis to first-order causal emergence, i.e., $\textrm{Un}(1)(V_{t}; X_{t} | X_{t}^{i})>0$ for all $i \in \{1, 2, \ldots, N\}$. Furthermore, we capitalise on a computationally efficient proxy measure for causal emergence, which was defined by Rosas and Mediano et al.~\cite{rosas_mediano_2020} as follows (here, only shown for $k$=1).

$$\psi_{t,t'}(V) = \textrm{I}(V_{t}; V_{t'})-\sum_{j}\textrm{I}(X_{t}^{i};V_{t'})$$  	 	

As can be seen, $\psi$ aims to capture the information that a supervenient feature $V_{t}$ provides about its own future, beyond what is already known after considering each of the system’s parts $X_{t}^{i}$ separately. Importantly, $\psi>0$ constitutes a sufficient criterion for $V_{t}$ to exhibit causal emergence. Note however, that $\psi<=0$ does not necessarily imply the absence of causal emergence. This is due to the fact that the system parts $X_{t}^{i}$ may share redundant information about the future of $V_{t}$, which is then subtracted multiple times. In fact, for this reason it is generally harder to detect causal emergence using $\psi$ in systems with high redundancy and many parts. 

It is worth pointing out that $\psi$ is always calculated for a specific supervenient feature $V_{t}$ and a time interval $\tau$ that determines the future time point $t'=t+\tau$ of the prediction target $V_{t'}$. In this study, $V_{t}$ represents the computational output of a RC. Furthermore, we chose $\tau=1$, which is the most meaningful choice in the context of one-step-ahead prediction as performed by the RCs. Finally, $\psi$ was computed using the software accompanying the theoretical work~\cite{rosas_mediano_2020}, available at https://github.com/pmediano/ReconcilingEmergences. 

\subsection{Human connectome data}

To construct bio-inspired RCs with human connectome topology, we used diffusion tensor imaging (DTI) data from the cohort of 100 unrelated, healthy subjects (46 males; 22-35 years) of the Human Connectome Project~\cite{vanEssen_2013}. The DTI scans were acquired in a 3T Siemens Skyra~\cite{ugurbil_2013} (acquisition scheme = monopolar gradient echo planar imaging (EPI); voxel size = 1.25 mm isotropic; repetition time (TR) = 5,500 ms; echo time (TE) = 89.50 ms; b-values = 1000, 2000, 3000 s/mm²; sampling directions = 90 directions per shell + 6 b0 images; flip angle = 2 apodised sinc RF pulses: 78°, 160°; 111 slices without gap).

A minimally preprocessed version~\cite{glasser_2013} of the data with corrections for eddy current distortions, susceptibility and motion artefacts was downloaded from the open-access HCP database. Subsequent preprocessing steps, as detailed by Luppi and Stamatakis~\cite{luppi_stamatakis_2020}, included transformation to MNI-152 standard space using the q-space diffeomorphic reconstruction tool, implemented in DSI Studio~\cite{yeh_2021_dsi-studio}, and a nonlinear registration algorithm from the statistical parametric mapping software (diffusion sampling length ratio = 2.5; output resolution = 1 mm). Finally, white matter fibres were reconstructed with FACT~\cite{yeh_2013_fact}, a deterministic tractography algorithm (angular cutoff = 55°; step size = 1.0 mm; minimum length = 10 mm; maximum length = 400 mm; spin density function smoothing = 0.0; the quantitative anisotropy threshold was determined by the signal in the cerebrospinal fluid). A white matter mask was generated by thresholding the resulting images at a quantitative anisotropy of 0.6 and reconstructed fibres that terminated outside of this mask were removed. The tractography algorithm was iterated until the maximum number of 1 million reconstructed streamlines was reached. The streamlines of neighbouring voxels were aggregated into 100 parcels representing brain regions according to the Schaefer parcellation~\cite{schaefer_2018}, resulting in a symmetric 100x100 matrix per subject encoding the number of axonal fibres between each pair of brain regions. Such a matrix is commonly termed ''connectome''~\cite{sporns_2005}.

Finally, we derived one group-level consensus connectome from the 100 subject-individual connectomes by averaging the streamline counts between each pair of brain regions across subjects and then setting all matrix entries, which were zero in more than half of the subjects, to zero. Bio-inspired RCs were constructed using the consensus human connectome.

\subsection{Reservoir computing}

\begin{table*}[t]
  \caption{RC hyperparameters. Hyperparameters marked with '*' were optimised.}
  \renewcommand\arraystretch{1}
  \def\rowsep{1pt}
  \small\centering
  \begin{tabular}{lll}
  
    \toprule
    
    {\bfseries Name} & {\bfseries Description} & {\bfseries Initialization} \\
    
    \midrule\midrule[.1em]
    
    \textit{N}
    & \makecell[l]{number of reservoir neurons}
    & \makecell[l]{100} \\
    [\rowsep]
    \hline

    \textit{C}
    & \makecell[l]{reservoir connection weights}
    & \makecell[l]{human connectome, or \\random weights $c_{i,j}\in[0;1]$ \\with $c_{i,j}=c_{j,i}$ and $c_{i,i}=0$} \\
    [\rowsep]
    \hline
    
    $\alpha*$
    & \makecell[l]{spectral radius of C}
    & \makecell[l]{$\alpha = \{0.1, 0.2, ..., 2.0\}$} \\
    [\rowsep]
    \hline
    
    $\beta*$
    & \makecell[l]{Tikhonov regularisation parameter}
    & \makecell[l]{$\beta = \{1.0, 1.5, ..., 1.0\}\textrm{x}10^{-8}$} \\
    [\rowsep]
    \hline

    $\rho*$
    & \makecell[l]{connection density of C, defined as \\$2E/(N(N-1))$, where $E$ are the \\ number of non-zero weights in C}
    & \makecell[l]{$\rho = \{0.01, 0.02, ..., 0.15\}$} \\
    [\rowsep]
    \hline

    $\sigma*$
    & \makecell[l]{maximum absolute input weight; \\ weights in $W_{in}$ were uniformly \\ sampled from $w_{in}\in[-\sigma; \sigma]$}
    & \makecell[l]{$\sigma = \{0.01, 0.02, ..., 0.1\}$} \\
    [\rowsep]
    \hline

    $\theta*$
    & \makecell[l]{input bias}
    & \makecell[l]{$\theta = \{0.1, 0.3, ..., 1.9\}$} \\

    \label{tab:hyperparameters}
  \end{tabular}
\end{table*}

\emph{Initialization}: The basic architecture of a RC in this study comprises i) a 3-dimensional input layer, ii) the reservoir, a network of 100 recurrently connected neurons, and iii) a 3-dimensional output layer. Input and output layers are fully and linearly connected to the reservoir via the weight matrices $W_{in}$ and $W_{out}$, respectively. The recurrent connections of the reservoir neurons are given by a symmetric matrix $C$, which is either a human connectome, or a symmetric, random matrix with uniform weights from 0 to 1. Upon initialization of each RC, the density of the reservoir network $C$ ($\rho$) was adjusted by setting the weakest weights in $C$ to zero. We chose to discard the weakest weights because tractography-based reconstructions of brain structural connectivity are prone to produce false positives, and stronger weights are more likely to represent true axonal fibres~\cite{maier-hein_2017}. After adjusting the network density, the weights of $C$ were scaled to a range from 0 to 1 and subsequently multiplied by the desired spectral radius ($\alpha$). Finally, all diagonal elements in $C$ were set to zero to improve the comparability between RCs with human connectome and random network topology as the human connectome lacks self-connections.

\emph{Workflow}: Following Platt et al.~\cite{platt_2021}, our reservoir computing paradigm involves three key operations: drive, train, and forecast. Firstly, during the drive operation, the reservoir dynamics are synchronised with that of the task environment by feeding an environmental input time series to the RC while updating the reservoir states $r_{t}^{N\textrm{x}1}$ according to 

\begin{equation}
    r_{t+1} = h\textrm{tanh}(Cr_{t}+W_{in}u_{t}+\theta)+(1-h)r_{t}
    \label{eq:reservoir-updates}
\end{equation}

where $u_{t}$ encodes the input at time $t$, $\theta$ is the input bias, tanh() denotes the hyperbolic tangent, and $h=0.005$ is the forward-Euler step size. Secondly, the train operation implements the computation of the RC output weights $W_{out}$ via Tikhonov ridge regression.

\begin{equation}
    W_{out} = u r^{\textrm{T}} (rr^{\textrm{T}} + \beta)-1
    \label{eq:wout-computation}
\end{equation}

Here, $\beta^{N\textrm{x}N}$ is a diagonal matrix with the Tikhonov regularisation parameter on its diagonal elements, and $u$ and $r$ respectively denote the input and reservoir-state time series that were generated during a preceding drive operation.

Thirdly, the forecast operation produces the forecast of the environmental dynamics in an iterative one-step-ahead prediction approach, where $u_{t}$ in equation~\ref{eq:reservoir-updates} is replaced by $W_{out}r_{t}$.

\begin{equation}
    r_{t+1} = h\textrm{tanh}(Cr_{t} + W_{in}W_{out}r_{t}+\theta)+(1-h)r_{t}
    \label{eq:forecasting}
\end{equation}

Building on these three operations, RCs were trained and evaluated as follows. First, the RC was driven with the initial 500 time points of the train input. This short drive operation, a.k.a. spinup~\cite{platt_2021}, serves to synchronise the RC with the environmental dynamics. Reservoir states generated during spinup are not used to compute $W_{out}$. Following spinup, the RC was driven with the remaining training input, and the reservoir states generated during this longer drive operation, which by default lasted 2000 time steps, were subsequently used to compute $W_{out}$. Next, the now trained RC was spun up (i.e. driven) with the first 500 time steps of the test input. This step may be seen as passing the initial condition of the prediction target to the RC. Finally, the environmental dynamics were forecast for 1000 time steps, and the prediction performance was evaluated.

\emph{Evaluation}: Inspired by Platt et al.~\cite{platt_2021}, we defined prediction loss as follows. Let $\epsilon_{i}(t)$ be the standardised error of predicting the dynamics of the $i$-th environmental variable:

\begin{equation}
    \epsilon_{i}(t):=|\hat{y}_{i}(t)-y_{i}(t)|\frac{1}{\sigma_{i}}
    \label{eq:prediction-error}
\end{equation}

where $|.|$ denotes the absolute difference, $\hat{y}_{i}(t)$ and $y_{i}(t)$ are the forecast and ground-truth trajectories of the environmental variable $i$ at time point $t$, respectively, and $\sigma_{i}$ is the standard deviation of $y_{i}(t)$ across time. Then, we define:

\begin{equation}
    \textrm{loss} := \frac{1}{TD}\sum^{T}_{t=1}\sum^{D}_{i=1}\epsilon_{i}(t)e^{\frac{-t}{T}}
    \label{eq:prediction-loss}
\end{equation}

where $T=1000$ is the length of the forecast, and $D=3$ is the number of environmental variables. Later error terms are given exponentially less weight to account for the exponential growth of prediction error due to the chaotic nature of the environment systems. We also estimated the emergence of each RC test output with respect to the reservoir states using $\psi$. Additionally, by thresholding loss and $\psi$ we derived two metrics with improved scalability and less sensitivity to outliers: the probability of emergence $\textrm{P}(E)$ where $E$ denotes the event of $\psi>0$, and the probability of prediction success $\textrm{P}(S)$ with $S$ denoting the event $\textrm{loss}<1$. By default, loss and $\psi$ values are reported as averages across 1 train and 100 test time series, and $\textrm{P}(E)$ and $\textrm{P}(S)$ are estimated based on the same 100 test time series.

\subsection{Hyperparameter optimization}

We optimised five RC hyperparameters: the spectral radius of the reservoir adjacency matrix C ($\alpha$), the density of C ($\rho$), the Tikhonov ridge regularisation parameter ($\beta$), input strength ($\sigma$), and input bias ($\theta$) (Tab.~\ref{tab:hyperparameters}). Hyperparameters were tuned for each type of reservoir network (human connectome or random) and for each task environment separately, using a microbial genetic algorithm~\cite{harvey_2011} and optimising either for performance or for emergence. When optimising for performance, the utility to be maximised, a.k.a. “fitness”, was given by negative loss or $\textrm{P}(S)$. When optimising for emergence, we maximised $\psi$ or $\textrm{P}(E)$ instead.

Genetic algorithms work by ''evolving'' a population of solutions with different hyperparameter combinations, encoded in so-called genotypes, towards high fitness through random mutations and competition between genotypes. We evolved populations of 100 RCs with different genotypes towards maximal prediction performance or emergence. Upon initialization, the hyperparameters of each RC in the population were sampled from a hyperparameter-specific discrete search space (Tab.~\ref{tab:hyperparameters}). The search spaces for $\beta$, $\sigma$, and $\theta$ were based on the respective hyperparameter choices in a similar study~\cite{platt_2021}. The search space for $\alpha$ was based on the widely accepted rule of thumb that RCs tend to transition from a stable to a chaotic regime at a spectral radius of about 1~\cite{yildiz_2012, suarez_2021}. Lastly, the search space for $\rho$ was constrained so as to not exceed the density of the consensus human connectome.

Prior to each evolutionary optimization, we generated 1 training and 100 test inputs from the given task environment, which provided the input data throughout the optimization. Subsequently, populations evolved for 3000 generations: First, the fitness of two randomly selected RCs in the population was compared. Next, each hyperparameter of the inferior genotype had a 20\% chance of being ''mutated'', i.e., shifted to the next higher or lower value in the corresponding search space. The direction of shift was chosen randomly if possible under the constraint that the new value remained within the search space. Following mutation, each hyperparameter of the inferior genotype was set to the corresponding hyperparameter value of the superior genotype with a probability of 0.2. 

\subsection{Task environments}

We implemented six 3-dimensional chaotic dynamical systems, which represented the task environments of the RCs: the Lorenz attractor~\cite{lorenz_1963}, and five Sprott chaotic flow systems~\cite{sprott_1994}. Sprott identified 19 algebraically simple systems with complex chaotic dynamics. From these, we selected systems A, B, G, K, and R due to their tendency to maintain values within manageable ranges. The ordinary differential equations (ODEs) describing the dynamics of each environment are listed in Tab.~\ref{tab:environments}. The environmental systems were simulated using forward Euler with a step size $h$=0.005 and $h$=0.05 for the Lorenz and Sprott systems, respectively. The initial condition of each generated time series was sampled by first drawing from uniform distribution over the approximate value range of each system variable (Tab.~\ref{tab:environments}), and then iterating the system for a random number of time steps between 1 and $\frac{1}{h}$. The resulting position of the system was used as the initial condition of the input time series to be generated.

\begin{table}[h!]
  \caption{ODEs of each task environment.}
  \small\centering
  \renewcommand\arraystretch{3}
  \def\rowsep{6pt}
  \begin{tabular}{lll}
  
    \toprule
    
    {\bfseries Environment} & {\bfseries ODEs} & {\bfseries Initialization} \\
    
    \midrule\midrule[.1em]
    
    \textit{Lorenz}
    & \makecell[l]{$dx/dt=10(y-x)$ \\$dy/dt=x(28-z)$ \\$dz/dt=xy-\frac{8}{3}z$}
    & \makecell[l]{$x\in[40;-20]$ \\$y\in[50;-25]$ \\$z\in[50;0]$} \\[\rowsep]
    \hline

    \textit{Sprott A}
    & \makecell[l]{$dx/dt=y$ \\$dy/dt=-x+yz$ \\$dz/dt=1-y^{2}$}
    & \makecell[l]{$x\in[10;-5]$ \\$y\in[10;-5]$ \\$z\in[10;-5]$}  \\[\rowsep]
    \hline
    
    \textit{Sprott B}
    & \makecell[l]{$dx/dt=yz$ \\$dy/dt=x-y$ \\$dz/dt=1-xy$}
    & \makecell[l]{$x\in[10;-5]$ \\$y\in[10;-5]$ \\$z\in[10;-5]$}  \\[\rowsep]
    \hline
    
    \textit{Sprott G}
    & \makecell[l]{$dx/dt=0.4x+z$ \\$dy/dt=xz-y$ \\$dz/dt=-x+y$}
    & \makecell[l]{$x\in[5;-3]$ \\$y\in[4;-3]$ \\$z\in[6;-3]$}  \\[\rowsep]
    \hline

    \textit{Sprott K}
    & \makecell[l]{$dx/dt=xy-z$ \\$dy/dt=x-y$ \\$dz/dt=x+0.3z$}
    & \makecell[l]{$x\in[6.6;-4.7]$ \\$y\in[4;-2.5]$ \\$z\in[6.7;-0.8]$}  \\[\rowsep]
    \hline

    \textit{Sprott R}
    & \makecell[l]{$dx/dt=0.9-y$ \\$dy/dt=0.4+z$ \\$dz/dt=xy-z$}
    & \makecell[l]{$x\in[7;-5]$ \\$y\in[7.5;-2.5]$ \\$z\in[10;-9]$}  \\

    \label{tab:environments}
  \end{tabular}
\end{table}

\subsection{Statistical analysis}

Statistical tests were performed using a permutation-based t-test and bootstrapping 10,000 times. Effect sizes were measured with Hedge's $g$. P-values were FDR-corrected for multiple comparisons.

\bibliography{refs}

\section{Appendix}

\end{document}